\documentclass[12pt]{article}
\usepackage{amsmath}
\usepackage{amsfonts}
\usepackage{amssymb}
\usepackage{amscd}
\usepackage{latexsym,eucal}
\usepackage{graphicx}
\begin{document}
%
\def\mkb{\mbox}
\def\beq{\begin{equation}}
\def\eeq{\end{equation}}
\def\beqn{\begin{eqnarray}}
\def\eeqn{\end{eqnarray}}
%
%
\newcommand{\mfootnote}[1]{{}}                                 
\newcommand{\mlabel}[1]{\label{#1}}                    
\newcommand{\mcite}[1]{\cite{#1}}                    
\newcommand{\mref}[1]{Eq. (\ref{#1})}

%
\newtheorem{theorem}{Theorem}[section]
\newtheorem{proposition}[theorem]{Proposition}
\newtheorem{definition}[theorem]{Definition}
\newtheorem{lemma}[theorem]{Lemma}
\newtheorem{coro}[theorem]{Corollary}
\newtheorem{prop-def}{Proposition-Definition}[section]
\newtheorem{claim}{Claim}[section]
\newtheorem{remark}[theorem]{Remark}
\newtheorem{example}[theorem]{Example}
\newtheorem{propprop}{Proposed Proposition}[section]
\newtheorem{conjecture}{Conjecture}
\newenvironment{exam}{\begin{example}\rm}{\end{example}}
\newenvironment{rmk}{\begin{remark}\rm}{\end{remark}}
%
%
%
%
\def\ta1{\includegraphics[scale=0.42]{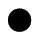}}
\def\tb2{\includegraphics[scale=0.42]{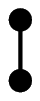}}
\def\tc3{\includegraphics[scale=0.42]{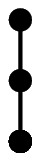}}
\def\td31{\!\!\includegraphics[scale=0.42]{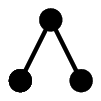}}
\def\te4{\includegraphics[scale=0.42]{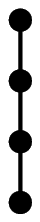}}
\def\tf41{\!\!\includegraphics[scale=0.42]{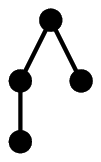}}
\def\tg42{\!\!\includegraphics[scale=0.42]{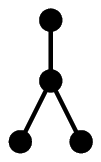}}
\def\th43{\!\!\includegraphics[scale=0.42]{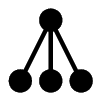}}
\def\ti5{\includegraphics[scale=0.42]{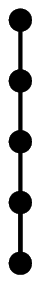}}
\def\tj51{\!\!\includegraphics[scale=0.42]{tree51}}
\def\tk52{\!\!\includegraphics[scale=0.42]{tree52}}
\def\tl53{\!\!\includegraphics[scale=0.42]{tree53}}
\def\tm54{\!\!\includegraphics[scale=0.42]{tree54}}
\def\tn55{\!\!\includegraphics[scale=0.42]{tree55}}
\def\tp56{\!\!\includegraphics[scale=0.42]{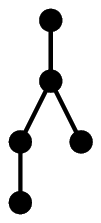}}
\def\tq57{\!\!\includegraphics[scale=0.42]{tree57}}
\def\tr58{\!\!\includegraphics[scale=0.42]{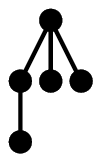}}
%
%
%
\def\sta1{\includegraphics[scale=0.62]{tree1}}
\def\stb2{\includegraphics[scale=0.62]{tree2}}
\def\stc3{\includegraphics[scale=0.62]{tree3}}
\def\std31{\!\!\includegraphics[scale=0.62]{tree31}}
\def\ste4{\includegraphics[scale=0.62]{tree4}}
\def\stf41{\!\!\includegraphics[scale=0.62]{tree41}}
\def\stg42{\!\!\includegraphics[scale=0.62]{tree42}}
\def\sth43{\!\!\includegraphics[scale=0.62]{tree43}}
\def\sti5{\includegraphics[scale=0.62]{tree5}}
\def\stj51{\!\!\includegraphics[scale=0.62]{tree51}}
\def\stk52{\!\!\includegraphics[scale=0.62]{tree52}}
\def\stl53{\!\!\includegraphics[scale=0.62]{tree53}}
\def\stm54{\!\!\includegraphics[scale=0.62]{tree54}}
\def\stn55{\!\!\includegraphics[scale=0.62]{tree55}}
\def\stp56{\!\!\includegraphics[scale=0.62]{tree56}}
\def\stq57{\!\!\includegraphics[scale=0.62]{tree57}}
\def\str58{\!\!\includegraphics[scale=0.62]{tree58}}
%
%
%
\def\A{\mathcal{A}}
\def\D{\mathcal{D}}
\def\F{\mathcal{F}}
\def\G{\mathcal{G}}
\def\H{\mathcal{H}}
\def\L{\mathcal{L}}
\def\M{\mathcal{M}}
\def\P{\mathcal{P}}
\def\RB{\mathcal{R}}
\def\T{\mathcal{T}}
\def\U{\mathcal{U}}
\def\X{\mathcal{X}}
\def\l{\mathfrak{l}}
\def\g{\mathfrak{g}}
\def\C{\mathbb{C}}
\def\K{\mathbb{K}}
\def\N{\mathbb{N}}
\def\R{\mathbb{R}}
\def\Z{\mathbb{Z}}
\def\to{\rightarrow}
\def\lto{\longrightarrow}
\def\sh{\sqcup \!\! \sqcup}
%
%
%
%
\title{Integrable Renormalization I: the Ladder Case}

{\small
\author{KURUSCH EBRAHIMI-FARD\footnote{kurusch@ihes.fr}\\
 I.H.\'E.S.\\ 
         Le Bois-Marie, 35, Route de Chartres\\
         F-91440 Bures-sur-Yvette, France\\
         {{and}}\\
         Universit\"at Bonn - 
         Physikalisches Institut\\ 
         Nussallee 12,
         D-53115 Bonn, 
         Germany\\
    \vspace{.3cm}\\
        LI GUO\footnote{liguo@newark.rutgers.edu}\\
        Rutgers University\\ 
         Department of Mathematics and Computer Science\\
         Newark, NJ 07102, USA\\
    \vspace{.3cm}\\
         DIRK KREIMER\footnote{kreimer@ihes.fr and dkreimer@bu.edu. Center for Math.~Phys., Boston University.}\\
         CNRS--I.H.\'E.S.\\
         Le Bois-Marie, 35, Route de Chartres\\ 
         F-91440 Bures-sur-Yvette, France\\
         }}
\maketitle
\vspace{-0.5cm} 
%
%
%
%
\begin{abstract}
In recent years a Hopf algebraic structure underlying the process of renormalization in 
quantum field theory was found. It led to a Birkhoff factorization for (regularized) 
Hopf algebra characters, i.e. for Feynman rules. 
In this work we would like to show that this Birkhoff factorization finds its natural 
formulation in terms of a classical r-matrix, coming from a Rota-Baxter structure underlying 
the target space of the regularized Hopf algebra characters.  
Working in the rooted tree Hopf algebra, the simple case of the Hopf subalgebra 
of ladder trees is treated in detail. The extension to the general case, i.e. 
the full Hopf algebra of rooted trees or Feynman graphs is briefly outlined.   
\end{abstract}
%
%
\section{Introduction}

In 1960 the mathematician Glen Baxter~\mcite{B} used a simple algebraic identity to solve 
an analytic problem. Later, in 1963 F. V. Atkinson \mcite{A} gave a characterization of 
this relation in terms of a so-called subdirect Birkhoff decomposition. It was G.-C. Rota who 
investigated this identity more thoroughly and realized its importance within 
combinatorics and other fields in mathematics \mcite{Rota1, Rota2, C}. 
This identity is called the Rota-Baxter relation and will be introduced in the next section. 
See \mcite{Rota3,LG2} for reviews and \mcite{LG1, KEF, PhL1, Agu1, AguLoday} for recent 
developments relating this identity to other fields in mathematics.

The same relation in its Lie algebraic version was later rediscovered under the name 
(operator form of the modified) classical Yang-Baxter\footnote{Here the relation is 
named after the physicists C.-N.Yang and Rodney Baxter.} equation within the field 
of integrable systems \mcite{BD, STS1, STS2}. See \mcite{STS3} for a nice review 
especially with respect to double Lie algebras and factorization theorems, 
i.e. the Riemann Hilbert problem.

Very recently the Rota-Baxter relation was found to be of crucial importance within the 
Hopf algebraic approach of Kreimer, and Connes and Kreimer to renormalization theory 
of perturbative quantum field theory \mcite{Kreim1, Kreim2, CK1, CK2}. It implies a  
Birkhoff decomposition of Hopf algebra characters with the target space being 
a Rota-Baxter algebra. See \mcite{Weinz1} for a first introduction.

Our work aims at a clarification of the link between the last two subjects. 
We show how the Rota-Baxter relation on the target space of the Hopf algebra characters can be 
lifted to the Lie algebra of infinitesimal characters. Atkinson's theorem then implies an 
infinitesimal factorization on this space. 
This factorization in turn implies the Birkhoff decomposition for the Hopf algebra characters. 
We derive its twisted antipode form first found by one of us \mcite{Kreim1}. There it was 
introduced in a formal way mainly motivated by its ability to reproduce Zimmermann's forest formula. 
Dimensional regularization together with a minimal subtraction scheme
furnishes one special case (distinguished by its convenience in applied particle physics)
of our approach in 
which the target space consists of the algebra of Laurent series with finite pole part. 
Its Rota-Baxter structure stems from the projection onto the pole part of a Laurent series.
Here, we will discuss the algebraic Birkhoff decomposition for general Rota--Baxter maps $R$,
provided by various choices of renormalization schemes.
Let us mention here that we will restrict our attention later on to the case of the Hopf subalgebra 
of rooted ladder trees. In this manner we can avoid lengthy formulas generated by the Baker--Campbell--Hausdorff
formula, and can treat this case in full technical detail.  
In the end it will be outlined though how the general case, i.e. for arbitrary rooted trees or graphs, 
is derived.  Details and also a mathematically more rigorous presentation will be given in a forthcoming paper,
emphasizing the link to integrable systems.

The paper is structured as follows.
The next section introduces the notion of a Rota-Baxter algebra and its main characterization by 
Atkinson's theorem. Section three contains the main results. We briefly introduce the Hopf algebra 
of rooted trees and the group of characters and its Lie algebra of infinitesimal characters, with 
target space of Rota-Baxter type. 
The Birkhoff decomposition is then derived. We stress that we obtain Bogoliubov's R-bar operation
from scratch, as a natural consequence of Atkinson's theorem.
We end this work with a summary and an outline describing the case when the Hopf algebra is not cocommutative.   
%
%
%
\section{Rota-Baxter algebras and the double construction}

Let $\K$ be a field of characteristic $0$. 
By a $\K$-algebra we mean an associative algebra over $\K$ that 
is not necessarily unital or commutative unless stated otherwise.  

\begin{definition} 
Let $\A$ be a $\K$-algebra with a $\K$-linear map 
$R: \A \to \A$. We call $\A$ a Rota-Baxter
$\K$-algebra and $R$ a Rota-Baxter map (of weight $\theta  \in \K$) if the operator $R$ 
holds the following Rota-Baxter relation of weight $\theta \in \K$
\footnote{Some authors denote this relation in the form 
$R(x)R(y)=R\big(R(x)y + xR(y)+\lambda xy\big)$. So $\lambda=-\theta$.}:
\beq
    R(x)R(y) + \theta R(xy) = R\big(R(x)y + xR(y)),\ \forall x,y \in \A. 
    \mlabel{RBR}
\eeq
\end{definition}
We note that a Rota-Baxter relation can be defined on $\A$ even when the multiplication on $\A$ 
is not associative, {\it e.g.} when $\A$ is a Lie or pre-Lie algebra. 
In the case $\theta=0$, the Rota-Baxter operator is somewhat degenerate (see Atkinson's theorem below).
When $\theta \ne 0$, a simple transformation $R \to \theta^{-1}R$ gives the standard
form of Eq. (\ref{RBR}). For the rest of the paper we will always assume the
Rota-Baxter map to be of weight $\theta=1$, i.e. to be in standard form.

\begin{rmk}
(1) If $R$ fulfills relation (\ref{RBR}) for $\theta=1$ then $\tilde{R}:=id-R$ fulfills the 
same Rota-Baxter relation.\\
(2) The images of $R$ and $id-R$ give subalgebras in $\A$.
\end{rmk}

\begin{exam}
(1) An important class of Rota-Baxter maps is given by certain projectors. 
This is the case for the minimal subtraction map $R_{MS}$ in renormalization theory, 
which is a Rota-Baxter map of weight $\theta=1$ on the algebra of Laurent series 
$\C\big[\!\big[\epsilon,\epsilon^{-1}]$ \cite{Kreim2}. 
For $\sum_{k\ge-m}^{\infty}c_k\epsilon^{k} \in \C\big[\!\big[\epsilon,\epsilon^{-1}]$:
\beq
    R_{ms} \big(\sum_{k \ge -m}^{\infty}c_k\epsilon^{k}\big):=\sum_{k\ge-m}^{-1}c_k\epsilon^{k}.
    \mlabel{Rms}
\eeq
(2) Another nice example \mcite{PhL1} of a Rota-Baxter map of weight $\theta \in \K$ is 
the operator $\beta: M^{up}_{n}(\K) \to M^{up}_{n}(\K)$ defined on the subalgebra of 
$n \times n$ upper triangular matrices $M^{up}_{n}(\K) \subset M_{n}(\K)$, mapping 
an element $x$ to the diagonal matrices 
$M^{up}_{n}(\K) \ni x \mapsto \beta(x) \in M^{d}_{n}(\K) \subset M^{up}_{n}(\K)$:
$$
  \big( \beta(x) \big)_{ij} := \delta_{ij} \theta \sum_{k \ge i}^{n} x_{ik}.
$$    
(3) The Riemann integral:
$$
 R[f](x):=\int_{0}^{x}f(y)dy
$$
provides an example for a Rota-Baxter map of weight zero; Eq. (\ref{RBR}) 
for $\theta=0$ gives the rule for integration by parts.
\end{exam}
We now introduce the modified Rota-Baxter relation, its Lie algebraic version can be found in \mcite{BD, STS1}.

\begin{definition}
Let $\A$ be a Rota-Baxter algebra, $R$ its Rota-Baxter map. 
Define the operator $B: \A \to \A,\;\; B:=id-2R$ to be the modified Rota-Baxter map and 
call the corresponding relation fulfilled by $B$:
\beq
    B(x)B(y) = B\big(B(x)y + xB(y)\big) - xy,\ \forall x,y \in \A
    \mlabel{mRBR}
\eeq
the modified Rota-Baxter relation. 
\end{definition}

\begin{proposition} 
In the case of the Rota-Baxter algebra $\A$ to be either an associative or pre-Lie 
$\K$-algebra, the (modified) Rota-Baxter relation naturally extends to the Lie 
algebra $\L_{\A}$ with bracket $[x,y]:=xy-yx,\; \forall x,y \in \A$:
\beqn
    \mlabel{LieRBR}  [R(x),R(y)]\! &\!\!+\!\!&\! R([x,y]) = R\big([R(x),y] + [x,R(y)]\big) \\ 
    \mlabel{mLieRBR} [B(x),B(y)]\! &\!\!=\!\!&\! B\big([B(x),y] + [x,B(y)]\big) - [x,y].
\eeqn
\end{proposition}
The proof of this follows from a simple calculation. 
The relations (\ref{LieRBR}) and (\ref{mLieRBR}) are well-known as the (operator form
of the) classical Yang-Baxter and modified Yang-Baxter equation.\\

The following Proposition~\ref{pp:double} and Theorem~\ref{thm:atk} characterize Rota-Baxter algebras. 

\begin{proposition} 
Let $\A$ be a Rota-Baxter algebra with (modified) Rota-Baxter map $R$ ($B=id-2R$).  
Equipped with the new product:
\beqn
  \mlabel{double}  a \ast_R b &:=& R(a)b + aR(b) - ab             \\ 
                              &=& -\frac{1}{2}\big(B(a)b+aB(b)\big),
\eeqn
$\A$ is again a Rota-Baxter algebra of the same type, denoted by $\A_R$. 
\mlabel{pp:double}
\end{proposition}
The proof of this proposition is immediate by the definition of $\ast_R $.
We call this new Rota-Baxter algebra $\A_R$ the double of $\A$.

\begin{rmk} 
(1) It is obvious that Proposition~\ref{pp:double} implies a whole hierarchy of doubles $\A^{(i)}_R$
(here, $\ast_R=\ast_R^{(1)} $):
$$
  \A^{(0)}_R:=\A,\; \A^{(1)}_R:=(\A, \ast_R), \; \cdots, \; \A^{(i)}_R:=(\A, \ast^{(i)}_R),\cdots
$$
$$
  a \ast^{(i)}_R b := \frac{d^i}{dt^i}_{|_{t=0}} e^{-\frac{1}{2}tB}(a)\: e^{-\frac{1}{2}tB}(b),\; a,b \in \A. 
$$
Let us call $\A^{(i)}_R$ the i-$th$ double of $\A$ and the double of $\A^{(i-1)}_R$.\\
(2) The Rota-Baxter map becomes an algebra homomorphism between $\A^{(i)}_R$ and 
$\A^{(i-1)}_R$, $i \in \N$:
\beq
  R(a \ast^{(i)}_R b)=R(a) \ast^{(i-1)}_R R(b). \nonumber
\eeq
\mlabel{rk:double}
(3) For the Rota-Baxter map $\tilde{R}:=id - R$, we have
\beq
  \tilde{R}(a \ast^{(i)}_R b)=-\tilde{R}(a) \ast^{(i-1)}_R \tilde{R}(b). \nonumber
\eeq
\end{rmk}
The last equation can be written as 
$\hat{R}:=-\tilde{R},\; \hat{R}(a \ast^{(i)}_R b) = \hat{R}(a) \ast^{(i-1)}_R \hat{R}(b)$.
As a side remark we should mention that the notion of the double of $\A$ for 
associative and non-associative algebras may be found in \mcite{STS1}.

A relation closely related to the Rota-Baxter relation (\ref{RBR}) is:
\beq
    N(x)N(y) + N^2(xy) = N\big(N(x)y + xN(y)\big), \;\; x,y \in \A.
    \mlabel{ANR}
\eeq
The map $N$ might be called an associative Niejnhuis operator or just
Nijenhuis map for short~\mcite{KEF2}. In this setting ``associative'' refers to
the relation (\ref{ANR}) to distinguish it clearly form its Lie
algebraic version \mcite{GS1, KSM}:
\beq
    [N(x),N(y)] + N^2([x,y]) = N\big([N(x),y] + [x,N(y)]\big).
    \mlabel{LNR}
\eeq
As in the case of the Rota-Baxter relation, a Niejnhuis map on a $\K$-algebra 
$\A$ also gives a Niejnhuis map for the associated Lie algebra 
$\L_{\A}:=(\A,[\:-\:,\:-\:])$, $[\:-\:,\:-\:]$ being the commutator. Also, similar to the case of 
the Rota-Baxter relation the associative Niejnhuis identity implies a hierarchy
of algebra products. 
We will not go further into details with respect to this relation.

\begin{proposition} 
Let $\A$ be a commutative, associative Rota-Baxter algebra. For $n \in \N, \; x \in \A$ we have 
\beqn
        (-R(x))^n &=& -R \big( x^n + \sum_{k=1}^{n-1}{n \choose k} (-R(x))^{(n-k)} \: x^k \big),         \mlabel{Rshuf1}\\ 
   \tilde{R}(x)^n &=& \tilde{R} \big( x^n + \sum_{k=1}^{n-1}{n \choose k} (-R(x))^{(n-k)} \: x^k \big). \mlabel{Rshuf2} 
\eeqn
%
\mlabel{pp:sha}
\end{proposition}
The proof works inductively. Proposition~\ref{pp:sha},
and equations (\ref{Rshuf1}, \ref{Rshuf2}) will lead us to the twisted antipode 
formula \mcite{Kreim1,Kreim2,CK1}.\\[0.2cm]

We come now to the important result of Atkinson, characterizing Rota-Baxter algebras.

\begin{theorem} 
{\rm (Atkinson \mcite{A}):} 
For a $\K$-algebra $\A$ with a linear map $R: \A \to \A$ to be a Rota-Baxter $\K$-algebra, 
it is necessary and sufficient that $\A$ has a subdirect Birkhoff decomposition. 
\mlabel{thm:atk}
\end{theorem}
It should be underlined here that this theorem is true quite generally, in the sense that the 
algebra needs not to be associative, nor commutative.\\

Essentially, the subdirect Birkhoff decomposition in this case means that the cartesian product 
$\D:=(R(\A),-\tilde{R}(\A)) \subset \A\times\A$ is a subalgebra in $\A \times \A$ and that every 
element $x \in \A$ has a unique decomposition $x=R(x)+\tilde{R}(x)$.\\
The double construction introduced here and Atkinson's theorem should be compared with the results
in \mcite{BD, STS1, STS2, STS3}.
%
%
%
\section{R-matrix approach to renormalization: the rooted ladder tree case}
We will now briefly introduce the Connes-Kreimer Hopf algebra of rooted trees \mcite{CK3, FGV, M}.  

\begin{definition} 
A rooted tree is a finite, connected oriented graph without loops in which  
every vertex has exactly one incoming edge, except one (the root) which has no incoming but only 
outgoing edges.
We denote the set of edges and vertices of a rooted tree $T$ by $E(T)$, $V(T)$ 
respectively.   
\end{definition}

Let us denote the set of (isomorphism classes of) rooted trees by $\T_{rt}$. 
$$
  \sta1 \;\;\;\; \stb2 \;\;\;\; \stc3 \;\;\;\; \std31 \;\;\;\; \stf41 \;\;\;\; \stg42 \;\;\;\; \sti5 \;\;\;\; 
  \stp56 \;\;\;\; \str58 \;\;\; \cdots
$$

Let $\H_{rt}$ be the commutative algebra generated by these symbols $T \in \T_{rt}$ 
(one for each isomorphism class). 
The commutative product $m_{\H_{rt}}: \H_{rt} \otimes \H_{rt} \to \H_{rt}$ is written as  concatenation  
$m_{\H_{rt}}(T',T'')=:T'T''$ and the empty tree is denoted by $1$ giving the unit. 
The algebra may be graded by the number of vertices $\#(T)=|V(T)|$ of the rooted tree $T$.
We equip this algebra with a counit map $\epsilon :\H_{rt} \to \K$, $\epsilon(1):=1$ and 
$\epsilon(T_1 \cdots T_n)=0$ for $T_1 \cdots T_n \ne 1$.

We now define the coproduct $\Delta: \H_{rt} \to \H_{rt} \otimes \H_{rt}$. 
For this we first introduce the notion of simple cuts on rooted trees. 
A simple or admissible cut $c_T$ of a tree $T$ is a subset of its edges such that along 
any path from its root to one of its leaves one meets at most one element of $c_T$.
Deleting the set $c_T \subset E(T)$ of edges in $T$ produces one tree $R_{c_T}$ still containing the
original root and a set of pruned rooted trees $P_{c_T}$, the roots of which are identified with the 
vertex which had the cut edge in $c_T$ as incoming edge. The following examples may be helpful in 
understanding the concept of simple cuts, $R_{c_T}$ and  $P_{c_T}$:
\begin{center}      
              \includegraphics[scale=0.62]{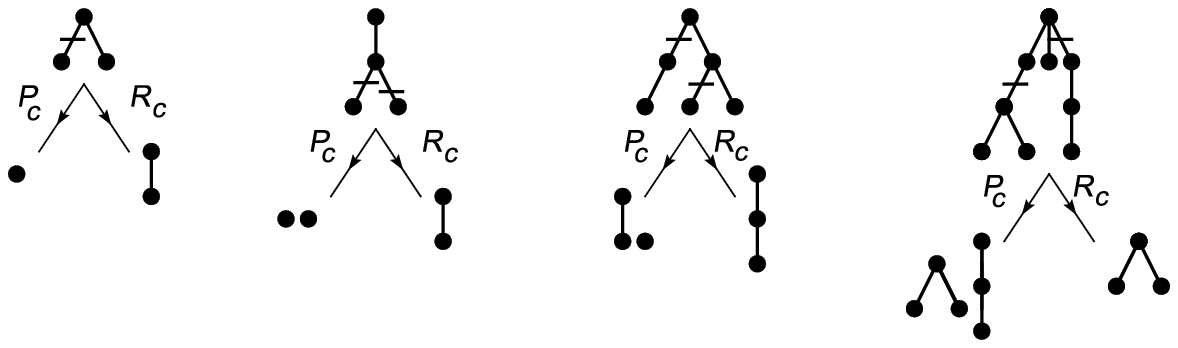}.
\end{center}
The coproduct may then be defined as follows. Let $C_T$ be the set of all admissible cuts of the 
rooted tree $T$. We exclude the empty cut $c^{(0)}_T,\; P_{c^{(0)}_T}= \emptyset,\; R_{c^{(0)}_T}=T$ 
and the full cut $c^{(1)}_T,\; P_{c^{(1)}_T}= T, \; R_{c^{(1)}_T}=\emptyset $:
\beq
    \Delta(T)= T \otimes 1 + 1 \otimes T + \sum_{c_T \in C_T} P_{c_{T}}(T') \otimes R_{c_T}(T''). 
    \mlabel{coprod}
\eeq
Here $T'$, respectively $T''$ stand for the rooted trees produced when applying $c_T$.
As an example we calculate the coproduct of the rooted tree \includegraphics[scale=0.3]{tree42} of 
weight $\#(\includegraphics[scale=0.22]{tree42})=4$:
\begin{center}      
              \includegraphics[scale=0.62]{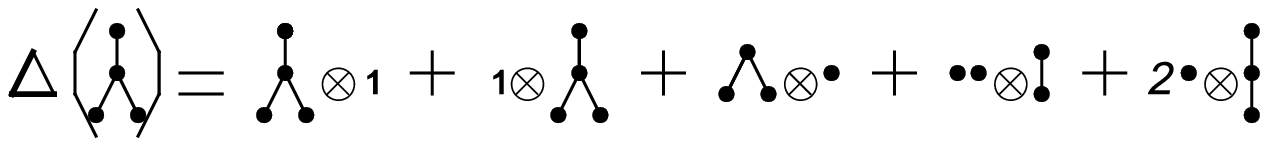}.
\end{center}
The coproduct (\ref{coprod}) is extended by linearity, and we define it to be an algebra homomorphism 
$\Delta(T_1 \cdots T_n +T_{n+1}\cdots T_m)=\prod_{i=1}^{n}\Delta(T_i) + \prod_{i=n+1}^{m}\Delta(T_i)$, 
$ \Delta(1):=1 \otimes 1$.
Obviously, (\ref{coprod}) is not cocommutative. So far we have a connected graded bialgebra, hence 
by general arguments a rooted tree Hopf algebra with antipode $S : \H_{rt} \to \H_{rt}$: 
\beq
    S(T):= -T - \sum_{c_T \in C_T} S(P_{c_{T}}(T'))R_{c_T}(T'').
    \mlabel{antipode}
\eeq
Again we exclude the empty and full cut, i.e. $c^{(0)}_T,\; c^{(1)}_T$ respectively, in the above sum.\\

The Hopf algebra $\H_{rt}$ contains a cocommutative Hopf subalgebra denoted by $\H_{l\:rt}$, 
generated by rooted ladder graphs:
$$
  \sta1 \;\;\;\;\;\; \stb2 \;\;\;\;\;\; \stc3 \;\;\;\;\;\; \ste4 \;\;\;\;\;\; \sti5 \;\;\;\;\;\; \dots,
$$
which we will denote in general by $t_n \in \H_{l\:rt} \subset \H_{rt}$, where $n \in \N$ is the number
of vertices, having at most one incoming and also at most one outgoing edge. By $t_0$ we denote the
unit, i.e. the empty tree $1$. The coproduct then becomes:
\beq
    \Delta(t_n)=t_n \otimes 1 + 1 \otimes t_n + \sum_{i=1}^{n-1}  t_i \otimes t_{n-i}. 
    \mlabel{ladcop}
\eeq
The graded dual $\H^*_{rt}$ equipped with the convolution product 
$f \star g  := m_{\K} \circ f \otimes g \circ \Delta,\; f,g \in \H^*_{rt}$ becomes an associative $\K$-algebra. 
We denote the pairing by brackets, $\langle f,T\rangle:=f(T) \in \K$.

Let $char_{\K}\H_{rt} \subset \H^*_{rt}$ be the group of characters , i.e. algebra morphisms into $\K$, 
with inverse $\phi^{-1}:=\phi \circ S$, $\phi \in char_{\K}\H_{rt}$. 
Let $\partial char_{\K}\H_{rt} \subset \H^*_{rt}$ be the Lie algebra of infinitesimal characters, i.e. 
derivations into $\K$:
\beq
    Z(T'T'') = Z(T')\epsilon(T'') + \epsilon(T')Z(T''), \; Z \in \partial char_{\K}\H_{rt}.
    \mlabel{infchar}
\eeq
with Lie bracket: 
\beq
    [Z',Z''] := Z' \star Z'' - Z'' \star Z'. \mlabel{Liechar}
\eeq 
$\partial char_{\K}\H_{rt}$ is generated by the infinitesimal characters $Z_T$ indexed by
rooted trees $T \in \T_{rt}$, and defined by:
\beq
    \langle Z_T,T'\rangle:=\delta_{T,T'}.
\eeq
Tree monomials are excluded from the index set due to the Leibniz rule (\ref{infchar}).
The Lie bracket for these generators is given by \mcite{CK3}:
\beq
    [Z_{T'},Z_{T''}] = \sum_{T \in \T_{rt}} \big( n(T',T'';T)-n(T'',T';T) \big)Z_T, \mlabel{Liebra}
\eeq 
where the $n(T',T'';T) \in \N$ denote so-called section coefficients which count the number 
of single simple cuts, $|c_T|=1$, such that $P_{c_{T}}=T'$ and $R_{c_T}=T''$: 
\beqn
\mlabel{cherry}
                 [Z_{\ta1},Z_{\tb2}]  &=& Z_{\tc3} + 2Z_{\td31} -  Z_{\tc3}  = 2Z_{\td31} \\ 
   \mlabel{ch}   [Z_{\td31},Z_{\ta1}] &=&  \frac{1}{2}[[Z_{\ta1},Z_{\tb2}],Z_{\ta1}]      \\  
                                      &=& Z_{\tg42} - 3Z_{\th43} - Z_{\tf41}.             \nonumber 
\eeqn
\begin{rmk} 
(1) The composition $Z_{T'} \star Z_{T''} = \sum_{T \in \T_{rt}} n(T',T'';T)Z_{T}$ defines a 
left pre-Lie algebra structure on $\partial char_{\K}\H_{rt}$.\\
(2) Antisymmetrizing this pre-Lie product gives the above Lie algebra, which lies at the heart 
of the combinatorics of renormalization theory in pQFT \mcite{CK4, KM}. 
\end{rmk}

The exponential map gives the bijection from $\partial char_{\K}\H_{rt} \to char_{\K}\H_{rt}$:
\beq
    exp^{\star}(Z):=\sum_{n=0}^{\infty} \frac{Z^{\star n}}{n!} \in char_{\K}\H_{rt}. \mlabel{exp}
\eeq
Where $Z$ is given as a series 
$$
  Z=\sum_{T \in \T_{rt}} a^T Z_T = a^{\ta1}Z_{\ta1} + a^{\tb2}Z_{\tb2} + a^{\tc3}Z_{\tc3} + a^{\td31}Z_{\td31} +\cdots
$$
The exponential map (\ref{exp}) is well defined since due to the Leibniz rule (\ref{infchar}) we have 
$Z^{\star n}(T)=0$ for $n >\#(T)$.\\

Following the Hopf algebraic approach to renormalization in perturbative quantum field theory (pQFT),
we introduce the notion of regularized (infinitesimal) characters, mapping $\H_{rt}$ into a 
commutative, associative, unital Rota-Baxter algebra $\A$\footnote{$\A=\C\big[\!\big[\epsilon,\epsilon^{-1}]$
in dimensional regularization together with minimal subtraction in pQFT, where the Rota-Baxter $R$ map is then given by 
equation~(\ref{Rms})}.
We therefore extend $\H^*_{rt}$ to $\H^*_{rt}\otimes \A = L(\H_{rt},\A)$, consisting of $\K$-linear maps from
$\H_{rt}$ into the Rota-Baxter algebra $\A$, i.e. 
$\langle \phi,T\rangle \in \A,\; \phi \in L(\H_{rt},\A),\; T \in \H_{rt}$.\\  

We then lift the Rota-Baxter map $R: \A \to \A$ to $L(\H_{rt},\A)$.

\begin{proposition} 
Define the linear map $\RB: L(\H_{rt},\A) \to L(\H_{rt},\A)$ by 
$f \mapsto \RB(f):=R \circ f: \H_{rt} \to R(\A)$. Then $L(\H_{rt},\A)$ becomes an associative, 
unital Rota-Baxter algebra.
The Lie algebra of infinitesimal characters $\L_{\H^*_{rt}} \subset L(\H_{rt},\A)$ with bracket 
(\ref{Liechar}) becomes a Lie Rota-Baxter algebra, i.e. for $Z',Z'' \in \partial char_{\A}\H_{rt}$,
\beq
    [\RB(Z'),\RB(Z'')]=\RB\big([Z',\RB(Z'')]\big) + \RB\big([\RB(Z'),Z'']\big) - \RB \big([Z',Z'']\big).
    \mlabel{Lie}
\eeq
\end{proposition}
Notice that we replaced $\K$ by $\A$ for the target space of the regularized infinitesimal characters.
The proof follows from the fact that $R$ is $\K$-linear and:
\beqn
     \RB(f)\star \RB(g)(T) &=& m_{\A} \big( \RB(f) \otimes \RB(g) \big) \circ \Delta(T) \nonumber \\ 
                         &=& m_{\A} \big( R(f(T_{(1)})) \otimes R(g(T_{(2)})) \big)     \nonumber \\  
                         &\stackrel{(\ref{RBR})}{=}& -R \circ m_{\A}\big( f(T_{(1)}) \otimes g(T_{(2)}) \big) + \nonumber \\
                         &&   \hspace{2cm} R \circ m_{\A}\big( f(T_{(1)}) \otimes R(g(T_{(2)})) \big) +         \nonumber \\
                         &&   \hspace{4cm} R \circ m_{\A}\big( R(f(T_{(1)})) \otimes g(T_{(2)}) \big)           \nonumber \\
                         &=& -R \circ m_{\A}\big( f \otimes g \big) \circ \Delta(T) +                           \nonumber \\
                         &&   \hspace{2cm} R \circ m_{\A}\big( f \otimes \RB(g) \big)\circ \Delta(T) +          \nonumber \\
                         &&   \hspace{4cm} R \circ m_{\A}\big( \RB(f) \otimes g \big)\circ \Delta(T)            \nonumber \\
                         &=& \RB(f \star \RB(g))(T) + \RB(\RB(f) \star g)(T) - \RB(f \star g)(T),
 \eeqn
where we used Sweedler's notation $\Delta(T)=\sum T_{(1)}\otimes T_{(2)}$ for the coproduct (\ref{coprod}), 
omiting the summation sign above.
For the second assertion, we only have to show that $\RB : \partial char_{\A}\H_{rt} \to \partial char_{\A}\H_{rt}$, 
but this again follows from the $\K$-linearity of $R$ and $\epsilon(T) \in \K$.

Using the double construction and Atkinson's theorem of section 2 we have the following\\[-0.3cm] 

\begin{lemma} 
The Rota-Baxter algebra $L(\H_{rt},\A)$ equipped with the convolution product:
\beq
     f \star_{\RB} g  =  f \star \RB(g) + \RB(f) \star g - f \star g
\eeq
gives a Rota-Baxter algebra structure on the set of linear functionals into the double $\A_R$ of $\A$, 
denoted by $L(\H_{rt},\A_R)$. An analog for $\L_{\H^*_{rt}}$ exists and is denoted by $\L_{\H^*_{rt}\:\RB}$.
$\RB$ becomes a (Lie) algebra morphism ($\L_{\H^*_{rt}\:\RB} \to \L_{\H^*_{rt}}$) 
$L(\H_{rt},\A_R) \to L(\H_{rt},\A)$.
\end{lemma} 

\begin{rmk} 
The above is also true for $\tilde{R}:=id-R$, respectively $\tilde{\RB}$ (see remark~\ref{rk:double}). We will
denote $\RB(\L_{\H^*_{rt}})$ by $\L^{-}_{\H^*_{rt}}$ and  $\tilde{\RB}(\L_{\H^*_{rt}})$ by $\L^{+}_{\H^*_{rt}}$.
\end{rmk} 
We now apply Atkinson's theorem to the Lie algebra $\L_{\H^*_{rt}}$ of infinitesimal characters, the
generators of the group of Hopf algebra characters $char_{\A}\H_{rt}$.\\[-0.3cm]

\begin{lemma} 
Every infinitesimal character $Z \in \L_{\H^*_{rt}}$ has a unique
subdirect Birkhoff decomposition $Z=\RB(Z) + \tilde{\RB}(Z)$.
\end{lemma} 

\begin{rmk}  
(1) In the case of an idempotent Rota-Baxter map $R$ we have a direct decomposition $\A=\A_- + \A_+$
respectively $\L_{\H^*_{rt}}=\L^-_{\H^*_{rt}} + \L^+_{\H^*_{rt}}$.\\
(2) Let $Z \in \L_{\H^*_{rt}}$ be the infinitesimal character generating the character 
$\phi=exp^{\star}(Z) \in char_{\A}\H_{rt}$. Using the result in proposition~\ref{pp:sha}, 
easily generalized to the non-commutative case, we then see that 
$exp^{\star}(\RB(Z))=\RB \big( exp^{\star_{\RB}}(Z) \big)$. 
\end{rmk}

Let us define $$b[\phi]:=exp^{\star_{\RB}}(Z)$$ which is a character of $\H_{rt} \to \A_R$, i.e.
$b[\phi] \in char_{\A_R}\H_{rt}$ and which we will call Bogoliubov's recursion $\bar{R}$-map for reasons which
will become clear soon.
Therefore 
\beqn
    exp^{\star}(\RB(Z))(T'T'') &=& \RB \big( exp^{\star_{\RB}}(Z) \big)(T'T'')        \nonumber \\
                               &=& R \big( b[\phi](T')\ast_R b[\phi](T'')\big)        \nonumber \\ 
                               &=&  exp^{\star}(\RB(Z))(T') exp^{\star}(\RB(Z))(T'').  \label{bogofact} 
\eeqn
We then have 
\begin{lemma} 
The Lie Rota-Baxter map $\RB$ $(\tilde{\RB})$ becomes a Lie group (anti-) homomorphism from 
$char_{\A_R}\H_{rt}$ to $char^{-(+)}_{\A}\H_{rt}$, where $char^{-(+)}_{\A}\H_{rt}$ are the 
Lie subgroups generated by the Lie subalgebras  $\L^{-(+)}_{\H^*_{rt}}$.
\end{lemma} 

As we already mentioned in the introduction, here we will only consider in detail the simple case of the
cocommutative Hopf subalgebra $\H_{l\:rt}$, respectively $L(\H_{l\:rt},\A)$.  
The latter is  generated by the $Z_{t_n} \in \partial char_{\A}\H_{rt}, \; n \in \N$ and necessarily is
an abelian Lie algebra $\L_{\H^*_{l\:rt}}$, $[Z_{t_n},Z_{t_m}]=0, \; n,m \in \N$.\\ 

The abelianess of $\L_{\H^*_{l\:rt}}$ and Atkinson's result imply the following theorem
(which extends to the non-abelian case using the appropriate BCH formulas
for the (multi-)commutator of $\RB(Z)$ with $\tilde{\RB}(Z)$).

\begin{theorem} 
{\rm (Ladder case of integral renormalization):}
Let $\phi \in char_{\A}\H_{l\:rt}$ be generated by $Z \in \L_{\H^*_{l\:rt}}$, i.e. $exp^{\star}(Z)=\phi$.
We have the following factorization: 
\beqn
     exp^{\star}(Z)=\phi &=& exp^{\star}\big(\RB(Z) + \tilde{\RB}(Z)\big)                    \nonumber \\
                         &=& exp^{\star}(\RB(Z)) \star exp^{\star}(\tilde{\RB}(Z)). \mlabel{ExpFac1}
\eeqn
\mlabel{thm:lad}
\end{theorem}
\begin{proposition} 
With the same assumption as in theorem~\ref{thm:lad} and the 
definitions $\phi^{-1}_{-}:=exp^{\star}(\RB(Z))$ and $\phi_{+}:=exp^{\star}(\tilde{\RB}(Z))$, we have
\beqn
     \phi_{-}(t_n) &=& exp^{\star}(-\RB(Z))(t_n)                                                              \nonumber    \\ 
                   &=& \RB\big(exp^{\star_{\RB}}(-Z)\big)(t_n)                                                \mlabel{A}   \\
                   &=& -\RB \big\{ \phi(t_n) + \sum_{k=1}^{n-1} \phi_{-}(t_k)\phi(t_{n-k}) \big\}             \mlabel{phi-}\\
     \phi_{+}(t_n) &=& exp^{\star}(\tilde{\RB}(Z))(t_n)                                                       \nonumber    \\                 
                   &=& -\tilde{\RB}\big(exp^{\star_{\RB}}(-Z)\big)(t_n)                                         \mlabel{B}   \\
                   &=& \tilde{\RB} \big\{ \phi(t_n) + \sum_{k=1}^{n-1} \phi_{-}(t_k)\phi(t_{n-k}) \big\}.     \mlabel{phi+} 
\eeqn
\mlabel{pp:phi}
\end{proposition}
The proof of this proposition follows immediately by proposition~\ref{pp:sha} and equations (\ref{Rshuf1}, \ref{Rshuf2}).
It can be generalized to the non-abelian case using the Hochschild cohomology of the Hopf algebra as in
\cite{Houches} and the resolution of the non-abelian Lie algebra in terms of its lower central series.

\begin{rmk} 
(1) From expressions (\ref{bogofact}, \ref{A}, \ref{B}) it is evident that $\phi_{\pm}$ are characters. We will see the 
same for the general case.\\
(2) For the ladder case we therefore arrive at the following result.
Since $exp^{\star}(-\RB(Z)) = -\RB(b[\phi])$ (remark 3.7) we have 
$$
  b[\phi](t_n) = exp^{\star_{\RB}}(Z)(t_n) = \phi(t_n) + \sum_{k=1}^{n-1} \phi_{-}(t_k)\phi(t_{n-k}).
$$
This justifies the name Bogoliubov's $\bar{R}$-map \mcite{CK1} for $exp^{\star_{\RB}}(Z)$, which finds its 
natural algebraic formulation as a character $exp^{\star_{\RB}}(Z) \in char_{\A_R}\H_{l\:rt}$ and which is mapped 
by the Rota-Baxter operator $\RB$ into $char_{\A}\H_{l\:rt}$. 
As mentioned before  this result carries over to the general case, i.e. to the 
non-cocommutative Hopf algebras of Feynman graphs or arbitrary decorated rooted trees.
Note that formulas (\ref{phi-}, \ref{phi+}) have been established already in \cite{Kreim1},
while to express (\ref{A}, \ref{B}) in a convenient way using the necessary BCH corrections will be reserved
to future work.
\end{rmk} 
Also, we would like to underline the similarity with the factorization theorems in \mcite{STS1, STS2, STS3}. 
We will dwell on this connection in greater depth in the future.

Let us summarize the result in proposition~\ref{pp:phi}. The (abelian) Lie algebra $\L_{\H^*_{l\:rt}}$ naturally extends
to a Lie Rota-Baxter algebra due to the Rota-Baxter structure on its target space. Therefore it contains two 
Lie algebra structures with respect to the original Lie bracket and the double coming from the Rota-Baxter map $\RB$.
Due to Atkinson's theorem it decomposes into two Lie subalgebras $\L^{-(+)}_{\H^*_{l\:rt}}$ which generate the  
Lie subgroups $char^{-(+)}_{\A}\H_{l\: rt}$. The infinitesimal decomposition on $\L_{\H^*_{l\:rt}}$ extends in the 
ladder case to the global factorization on the Lie group $char_{\A}\H_{l\:rt}$. We have the following diagrams on 
the Lie algebra level, respectively Lie group level. 

Let $b$ denote Bogoliubov's recursion formula, which is defined in terms of the exponential with respect to the double
product $\star_{\RB}$. We define $\G^l:= char_{\A}\H_{l\:rt}$, $\G^l_{\RB}:=char_{\A_R}\H_{l\:rt}$ and 
$\G^{l\:\pm}:=char^{\pm}_{\A}\H_{l\:rt}:$
\beq
    \L_{\H^*_{l\:rt}\:\RB} \xrightarrow{(\RB, \RB-id)} \big(\L^{-}_{\H^*_{l\:rt}},\L^{+}_{\H^*_{l\:rt}}\big) 
                        \xrightarrow{(id, - id)}  \L_{\H^*_{rt}}.  
\eeq
Here, $Z=\RB(Z)-(\RB-id)(Z)=(id, -id)\circ(\RB, \RB-id)(Z)$ gives the infinitesimal factorization. i.e. 
Atkinson's theorem.
\beq
   \G^l  \xrightarrow{b}  \G^l_{\RB}  \xrightarrow{(-\RB \otimes \tilde{\RB})} (\G^{l\:-},\G^{l\:+})
                                      \xrightarrow{m_{\G^l}}  \G^l.  
\eeq
The last diagram contains the global factorization, i.e. on the level of the Lie group $\G^l$ coming form the 
Lie algebra $\L_{\H^*_{l\:rt}}$.\\[-0.3cm]

\begin{rmk} 
(1) Using the idea of normal coordinates in \mcite{Mex1} we may relate the simple rooted ladder graphs,  
given by Schur polynomials $P^{(n)}(t_1,\dots,t_n)$\footnote{Schur polynomial of order $n$, $P^{(n)}$: Taylor expansion of 
$log \bigg( \sum_{n\ge0} t_{n} x^n \bigg)= \sum_{n\ge0} P^{(n)}(t_1,\dots,t_n) x^n \mcite{Mex1}$} of order $n$ 
for each rooted ladder tree $t_n$ and the $\phi_{\pm}$ character in the following way.
Starting with the regularized character $\phi :\H_{rt} \to \A$, we define the series (we omit tensor product signs)
$$
  Z_{\phi}:=\sum_{n>0} Z_{t_n}\phi(P^{(n)}(t_1,\dots,t_n)) \in char_{\K}\H_{rt}\otimes \A.  
$$
It then follows from \mcite{Mex1} that $exp^{\star}(Z_{\phi})(t_n)=\phi(t_n)$, i.e. $Z_{\phi}$ is the 
infinitesimal character generating the Lie group element $\phi$.  
As an example we calculate $\phi_{-}\big(\tc3\big)$:
\beqn
     exp^{\star}(-\RB(Z_{\phi}))\big(\tc3\big) &=&\bigg\{ -Z_{\tc3}R\big(\phi(P^{(3)}(\ta1,\tb2,\tc3))\big)                            \nonumber \\
                                     & & + \frac{1}{2}\big(Z_{\ta1}\star Z_{\tb2} + Z_{\tb2}\star Z_{\ta1}\big) 
                                                                                     R(\phi(\ta1))R\big(\phi(P^{(2)}(\ta1,\tb2))\big)     \nonumber \\
                                     & & + \frac{-1}{3!}Z_{\ta1} \star Z_{\ta1} \star Z_{\ta1} R(\phi(\ta1))^{3}\bigg\}\bigg(\tc3\bigg)   \nonumber \\   
                                     &=& -R\bigg(\phi\big(P^{(3)}(\ta1,\tb2,\tc3)\big) - R(\phi(\ta1))R\big(\phi(P^{(2)}(\ta1,\tb2))\big) \nonumber\\
                                     &&                     \hspace{3cm}        +\frac{1}{3!}R(\phi(\ta1))^{3} \bigg).       \nonumber 
\eeqn
(2) Let us briefly outline the approach to the general case. The full Lie algebra of infinitesimal characters
$\L_{\H^*_{rt}}$ is of course non-abelian and therefore the factorization has to include contributions 
in a subtractive manner to eliminate BCH terms. This may be achieved in a systematic way using the 
Baker-Campbell-Hausdorff (BCH) functional
$$
  BCH(A,B):=\frac{1}{2}[A,B] + \frac{1}{12}\big( [A,[A,B]]-[B,[A,B]]\big) + \cdots
$$
which starts with a commutator, i.e. is of order $>1$ in the number of infinitesimal characters. Recursively, order
by order in the grading on $\H_{rt}$, the correct contributions may be calculated. The first term of second order
essentially takes care of the cherry tree $\;\td31$ renormalization:
$$
  \chi^{(2)}=-\frac{1}{2}[\RB(Z),Z]
$$
for $\phi=exp^{\star}(Z) \in char_{\A}\H_{rt}$ and which should be compared to (\ref{cherry}).
Again, the normal coordinates theorem in \mcite{Mex1} provides a convenient way
to identify terms from the BCH formula.
This way we derive the implicitly given formulae for $\phi_{\pm}$ in \mcite{CK1}, respectively 
give an explicit formula for them in terms of the exponential map and an element in the image 
of the Rota-Baxter maps $\RB$, respectively $\tilde{\RB}$.\\
(3) The presented setting resembles the loop algebra-group case of integrable systems theory. 
The generalization takes place by using a general Rota-Baxter algebra as the target space.  
In a later publication we will dwell more carefully on this point.\\
(4) Following the recent work by Sakakibara \cite{Jap} we derive the scattering type formula 
for $\phi_{\pm}$. We first extend the Lie algebra $\partial char_{\A}\H_{rt}$ by an element $Z_{0}$ 
such that $[Z_{0},Z_{T}]=Y(Z_{T}):=\#(T)Z_{T}$ where $Y$ is the grading operator $Y(T):=\#(T)$, i.e. 
a derivation on $\H_{rt}$ (see \cite{CK2} for details). This implies an one parameter group 
$\theta_t \in Aut(\H_{rt})$ acting on $char_{\A}\H_{rt}$ by 
$$
  \langle\theta_t(\phi),T\rangle:=\langle\phi,\theta_t(T)\rangle, 
$$
i.e. $\theta_t = Ad_{ exp^{\star}(tZ_0)}$, and such that $\frac{d\theta_t}{dt}|_{t=0}=Y$. We then define 
a so-called $\beta$-function: 
\beqn
     \beta(\phi)&:=&\phi_{\pm}\star Y(\phi^{-1}_{\pm})  \nonumber \\
                &=&\phi_{\pm}\star [Z_{0},\phi^{-1}_{\pm}]=\phi_{\pm}\star Z_{0}\star \phi_{\pm}^{-1} - Z_0, \nonumber
\eeqn 
such that \cite{Jap} 
\beqn
     \exp^{\star}\big(t( \beta(\phi) + Z_0) \big)\star exp^{\star}(-tZ_0) &=& \phi_{\pm} \star \theta_t(\phi^{-1}_{\pm})   \nonumber \\
                                                                         &\xrightarrow{t\to\infty}& \phi_{\pm}.       \nonumber
\eeqn 
This should be compared to the expression we found for $\phi_{\pm}$ in terms of the Rota-Baxter map $\RB$.
\end{rmk}
%
%
\section{Summary, Conclusion and Outlook}
In earlier work \cite{CK1}  the combinatorics of renormalization in pQFT was described in terms of a 
Birkhoff factorization of the regularized Hopf algebra characters. The identification
of the singular part $\phi_{-}$ as a character relies on the Rota-Baxter structure
on the target space of the characters. Emphasizing the latter point and restricting 
for pedagogical reasons to the simple case of rooted ladder trees we were able to derive the twisted 
antipode formula for the singular part of the Birkhoff decomposition, solely using the 
properties of the Rota-Baxter map lifted to the Lie algebra of infinitesimal characters.

Extending this simple exercise to the general case of a non-cocommutative Hopf algebra
poses no conceptual challenge but becomes technically more demanding and will be presented elsewhere.

In our view it is important to underline  that it is the Lie algebra of rooted trees, or 
more generally the insertion-elimination Feynman Lie algebra, which completely  
describes the process of renormalization in pQFT. This will become even more apparent
when we treat the general case, indicated in the last remark in section 3.
These results strongly indicate that there is a deeper link between the realm of (classically)
integrable systems and the RG-flows. The last point especially demands for a deeper understanding 
of the Lie algebra of infinitesimal characters, respectively the Lie Group of regularized 
characters. We hope that this may be partly achieved by analyzing the link to the well 
established field of integrable systems.       
%
%
\section*{Acknowledgements}
The first author would like to thank the Ev. Studienwerk for financial support. Also the I.H.\'{E}.S.
and its warm hospitality is greatly acknowledged. 
%
%
%
%
%

%
\end{document}